\documentclass[%
12pt,T
 oneocolumn,
 nofootinbib,
 amsmath,amssymb,
 aps,
floatfix,
unsortedaddress
]{revtex4-2}

\usepackage{graphicx,color}
\usepackage{subfig}
\usepackage{dcolumn}
\usepackage{bm}
\usepackage{graphicx,color}
\usepackage{float}
\usepackage{multirow}
\usepackage{amsmath}
\usepackage{mathtools}
\usepackage{cancel}
\usepackage[pdfpagelabels, pdfencoding=auto, psdextra]{hyperref}
\hypersetup{%
 pdfsubject=Paper,
 pdfkeywords={nuclear physics} {Pionless EFT} {Two-Nucleon} {Lattice} {L\"usher formula},
 unicode = true,
 breaklinks = true,
 colorlinks = true,
 linkcolor = blue,
 citecolor = blue,
 menucolor = blue,
 citecolor = blue,
 urlcolor = blue
}

\graphicspath{{./}}
\begin{document}

\title{	Bound states of $ _{c\bar{c}}^{9}$Be within $c\bar{c}+\alpha+\alpha$ cluster models based on state-of-the-art QCD charmonium-nucleon interactions}
\author{Faisal Etminan}
 \email{fetminan@birjand.ac.ir}
\affiliation{
 Department of Physics, Faculty of Sciences, University of Birjand, Birjand 97175-615, Iran
}%
\affiliation{ Interdisciplinary Theoretical and Mathematical Sciences Program (iTHEMS), RIKEN, Wako 351-0198, Japan}

\date{\today}%
\begin{abstract}
	The possible bound state of the $ _{c\bar{c}}^{9}$Be, a charmonium-nucleus system, is investigated. The analysis is carried out within a three-cluster model, where its binary subsystems are represented as $ c\bar{c}\textrm{+}\alpha $ and $\alpha+\alpha$. The hyperspherical harmonics method is employed to facilitate a convenient description of this three-cluster configuration. 
The calculations are done by employing the effective $ c\bar{c}\textrm{-}\alpha $ potentials.
 These potentials were derived recently based on state-of-the-art lattice QCD calculations by HAL Collaboration, which provided interactions for 
the spin 3/2 $J/\psi N $, spin 1/2 $J/\psi N $, spin 1/2 $\eta_{c}N$ and spin-averaged $J/\psi N$ interactions, all obtained with nearly physical pion masses. The Coulomb interaction was also incorporated into the current calculations.
It is determined that, despite neither the $ _{c\bar{c}}^{5}$He nor the $^{8}$Be binary subsystems being bound, a bound state of the $ c\bar{c}\textrm{-} \alpha\alpha$ nuclear system could potentially exist. The maximum central binding energy is found to be approximately 1.71 MeV, based on the spin 1/2 $J/\psi N $ interaction, while a minimum value of about 0.56 MeV is obtained from calculations involving the spin 1/2 $\eta_{c}N$ interaction.
\end{abstract}


\maketitle
\section{Introduction} \label{sec:intro}
In addition to numerous motivations for investigating
the charmonium states such as $J/\psi$ and $\eta_{c}$, the interactions of these states with atomic nuclei are considered to provide valuable insights into the mechanisms of strongly interacting matter and the workings of quantum chromodynamics (QCD). Given that charmonia ($c\bar{c}$) and nucleons predominantly do not share light quarks, their interactions mediated through meson exchange involving light quarks are suppressed by the OZI rule. Should such states be bound within nuclei, identifying alternative attraction sources—such as gluon exchange—becomes crucial, as these could facilitate the binding of charmonia to atomic nuclei. 

The hypothesis proposed by Brodsky~\cite{BrodskyPRL1990}—that QCD van der Waals forces generated by multiple gluon exchanges could produce a binding energy of approximately 400 MeV in an $A=9$ nucleus—has spurred extensive research. This research aims to explore mechanisms beyond meson exchange, with considerable efforts devoted to investigating the potential existence of exotic bound states~\cite{KREIN2018161}. The original proposition suggested that QCD van der Waals forces might bind a charmonium state within a nucleus through an attractive potential. This was initially estimated via a variational approach utilizing a phenomenological Yukawa potential for the charmonium–nucleus interaction. Taking into account the nucleon distribution within the nucleus by folding the Yukawa potential with the nuclear density, later studies~\cite{WassonPRL1991} predicted a maximum binding energy of about 30 MeV in large nuclei.

Further investigations have employed various approaches—including QCD sum rules~\cite{KIM2001517,PhysRevC.82.045207}, charmonium color polarizability~\cite{PhysRevD.71.076005}, one-boson exchange models~\cite{PhysRevC.63.044906}, effective Lagrangians~\cite{PhysRevC.75.064903}, the quark model~\cite{PhysRevC.75.064907}, and for the first time, lattice QCD simulations to probe low-energy charmonium-hadron interactions~\cite{PhysRevD.74.034504}.

Lattice QCD simulations detailed in Ref.~\cite{Beane2015} demonstrated the existence of quarkonium-nucleus bound states for systems with fewer than five nucleons, with a binding energy in the vicinity of 60 MeV. It should be noted that these calculations were performed at the $SU(3)$ flavor symmetry point with unphysical pion masses, approximately $m_\pi \approx   805$ MeV.

Additionally, the energies associated with $\eta_{c}$–nucleus bound states are evaluated for various nuclei in Ref.~\cite{COBOSMARTINEZ2020135882}, whereby crucial inputs—including medium-modified masses of $D$ and $D^{*}$ mesons and nuclear density distributions—are derived from the quark-meson coupling model.

Due to recent theoretical advancements in HAL QCD methodologies~\cite{ishii2012, aoki2013, sasaki2020, etminan2024prd} and the increasing availability of high-performance computing resources, hadron-hadron interactions have been successfully extracted from first-principles lattice QCD simulations at nearly physical quark masses. These studies encompass a variety of interactions, including $\Lambda\Lambda$, $\Xi N$~\cite{sasaki2020}, $\Omega N$~\cite{Iritani2019prb}, $\Omega \Omega$~\cite{Gongyo2018}, $\Omega_{ccc}\Omega_{ccc}$~\cite{yanPrl2021}, $\phi$N~\cite{yan2022prd}, and $c\bar{c}$-N~\cite{LyuPLB2025}. Notably, the work presented in Ref.~\cite{LyuPLB2025} offers a comprehensive lattice QCD investigation into low-energy interactions between charmonium states and nucleons, including spin-$\frac{3}{2}$ $J/\psi N$, spin-$\frac{1}{2}$ $J/\psi N$, and spin-$\frac{1}{2}$ $\eta_c N$. These interactions demonstrate an overall attractive nature across all distance scales.

Although a consensus exists among theorists regarding the likelihood of charmonium-nucleus bound states, predictions for their binding energies display considerable variation, and these states remain experimentally unobserved despite numerous efforts. Recent measurements have begun to shift this perspective, notably the JLab GlueX Collaboration's~\cite{PhysRevLett.123.072001} measurement of the total $J/\psi$ photoproduction cross section near threshold. The prospects for further advancements are promising with ongoing and upcoming experiments at JLab~\cite{near2012psi, meziani2016}. Notably, there is a proposal to measure $J/\psi$ photoproduction off deuterons~\cite{baker2018study}.

Recently, femtoscopy—a technique that analyzes correlations between particles emitted in high-energy collisions—has emerged as a powerful alternative for probing strong interactions, surpassing traditional scattering experiments~\cite{Fabbietti:2020bfg, LIU20251}. In this context, correlation functions in the charm sector have been measured, including those involving $D^{-}p$~\cite{PhysRevD.106.052010}, $D\pi$, and $DK$ pairs~\cite{PhysRevD.110.032004}. 
These channels are nearly inaccessible through conventional scattering experiments, but valuable information has been obtained through such measurements, complemented by theoretical analyses~\cite{BRODSKY1997125, WU2025, Liang-ZhenPRD2025, liu2025charmoniumnucleonfemto}.
Specifically, in a series of works~\cite{krein2020femtoscopy, Krein2022EPJ, krein2023femtoscopy}, Krein et al. suggested the use of femtoscopic measurements of $J/\psi$-proton correlation functions in high-energy hadron collisions to access information about the low-energy $J/\psi$-nucleon interaction and  the origin of the proton’s mass.

In this work, the hyperspherical harmonics (HH) expansion method~\cite{Zhukov93,Casal2020,ETMINAN2023122639} is employed, and it is assumed that the $ _{c\bar{c}}^{9}$Be nucleus ($ c\bar{c}\textrm{-} \alpha\alpha$ system), composed of a $c\bar{c}=J/\psi, \eta_{c}$ and two $\alpha$ clusters, can be modeled as a three-cluster system. The search is conducted for possible bound states within the $ c\bar{c}\textrm{-} \alpha\alpha$ system, along with the determination of its ground state properties, such as matter radii.

The structure of this paper is organized as follows: Section~\ref{sec:Two-body-potentials} introduces the binary potentials, namely $c\bar{c}\textrm{-}\alpha$ and $\alpha\alpha$ interactions. A concise overview of the three-body hyperspherical basis formalism is provided in Section~\ref{sec:Three-body hyperspherical}. The results are presented and analyzed in Section~\ref{result}. Finally, Section~\ref{sec:Summary-and-conclusions} offers a summary and concluding remarks.

\section{Binary potentials }
\label{sec:Two-body-potentials}
{\textit{$ c\bar{c} $-N interactions and single-folding $ c\bar{c}\textrm{-}\alpha $ potential}- } \label{sec:folding-Model}
A realistic lattice QCD simulation of the $S$-wave $c\bar{c}$-N potentials was presented by the HAL QCD Collaboration, focusing on the channels $J/\psi N$ ($^{4}S_{3/2}$), $J/\psi N$ ($^{2}S_{1/2}$), and $\eta_c N$ ($^{2}S_{1/2}$), evaluated near the physical pion mass $m_{\pi} = 146.4(4)$ MeV~\cite{LyuPLB2025}. The notation $^{2s+1}L_{J}$ is employed, where $s$ denotes total spin, and $L$ and $J$ correspond to orbital and total angular momentum quantum numbers, respectively. 
Configurations of $(2+1)$-flavor gauge fields were generated on a large lattice volume of approximately $(8.1\,\textrm{fm})^3$, with simulations conducted at the imaginary-time slice $t/a=14$, where the lattice spacing $a=0.0846$ fm. The isospin-averaged hadron masses, accompanied by their statistical uncertainties, are recorded as follows: $K$ at $524.7(2)$ MeV, $N$ at $954.0(2.9)$ MeV, $J/\psi$ at $3093.5(1)$ MeV, and $\eta_c$ at $2994.2(1)$ MeV.

The interactions, derived from spacetime correlation functions of the charmonium–nucleon system via the HAL QCD method, are characterized by an attractive nature across all distances and feature a long-range tail indicative of a two-pion exchange potential. It has been reported in Ref.~\cite{LyuPLB2025} that the $J/\psi N$ potential exhibits a marginally greater strength at short range compared to the $\eta_c N$ potential. Conversely, the potentials for $J/\psi N$ ($^{4}S_{3/2}$), $J/\psi N$ ($^{2}S_{1/2}$), and $\eta_c N$ ($^{2}S_{1/2}$) display similar behaviors at long distances and approach near degeneracy, suggesting a common underlying mechanism governs these interactions at large separations.

As noted in the Introduction, the low-energy interaction between a nucleon and a heavy quarkonium ($Q\bar{Q}$) is believed to be predominantly mediated by multiple-gluon exchange; however, gluons are unable to propagate over large distances due to color confinement. Consequently, the dominant degrees of freedom at such distances involve color-neutral states coupled to gluons, with the lightest two-pion state playing a significant role~\cite{Brambilla2016}. 
It has been explicitly demonstrated by Lyu et al. in Ref.~\cite{LyuPLB2025} that their lattice results support the long-range $c\bar{c}$–N interaction being consistent with a two-pion exchange (TPE) mechanism for $Q\bar{Q}$–N interactions~\cite{PhysRevD.98.014029}.

Given the strong binding nature of the $\alpha$-particle and its low likelihood for property variations, an effective potential for the $c\bar{c}\textrm{-}\alpha$ system was estimated using the single-folding method, which integrates the nucleon density distribution within the $\alpha$-particle with the HAL QCD $c\bar{c}$–N interactions. For practical purposes aligned with the Dover-Gal potential model~\cite{dover1983}, the resulting $c\bar{c}\textrm{-}\alpha$ potential was fitted to an analytical Woods-Saxon (WS) form:
\begin{equation}
U_{c\bar{c}\textrm{-}\alpha}\left(r\right)=-\frac{U_{0}}{1+\exp\left(\frac{r-R}{t}\right)} , \label{eq:ws-fit}
\end{equation}
where $U_0$ represents the depth, $R$ is the nuclear radius, and $t$ signifies the surface diffuseness. These parameters were determined within the region $r \gtrsim 1.9$ fm, corresponding to the experimental root-mean-square radius of the alpha particle~\cite{filikhin2024folding}. The parameters, taken directly from Ref.~\cite{etminan2025alphaCcbar}, are listed in Table~\ref{tab:Charm-alpha-para} for different $c\bar{c}$–N spin channels.

When the distances between \(\alpha\) and \(c\bar{c}\) are sufficiently large, the clustering can be described as \(c\bar{c} + (NNNN)\). However, inside and near the \(\alpha\) cluster, all possible five-body configurations of the clustering should be considered. 
Since no bound states are formed by the \(c\bar{c}\) meson in any subsystem, it is assumed that the \(c\bar{c} + (NNNN)\) clusterization is dominant, and therefore the folding potential is regarded as an applicable approach for the \(c\bar{c}\textrm{-}\alpha\) interaction~\cite{filikhin2024phihe}.  
Thanks to the three-Gaussian analytical fitting of the HAL QCD potential, the folding integral can be solved analytically for the Gaussian matter distribution of the \(\alpha\) cluster (see the Appendix of Ref.~\cite{filikhin2024phihe}).

The single folding potential is found to be sensitive to the root-mean-square (rms) radius of \({}^{4}\)He. Recently, the rms matter radius of \({}^{4}\)He has been measured to be \(1.70 \pm 0.14\) fm~\cite{PhysRevC.109.L012201}. Given that the accurate measurement of the radius represents a significant puzzle, densities reproducing the rms radii \(1.70 \pm 0.14\) fm are employed in the folding potential approach. The influence of the rms radius on the obtained \(\phi \alpha\) potential has been investigated in Refs.~\cite{filikhin2024phihe, ETMINAN2025PLB139564}.

The folding procedure has been successfully applied in Refs.~\cite{Filikhin_2005, HiyamaPRC2022, filikhin2024phihe, etminanPRCphiNalpha2025} for analyses of the \(\Lambda NN\alpha, \Xi N\alpha\alpha, \phi\alpha\alpha, \phi N\alpha\) systems, and in Ref.~\cite{filikhin2000alpha8be}, the first \(0_{2}^{+}\) excited state of \({}^{12}\)C is described on the basis of the Faddeev equation. Agreement with experimental data and with calculations conducted within various theoretical frameworks has been testified by the predictions of these works.

Using these fitted potentials, the Schrödinger equation was solved to extract binding energies. It was found that only the attractive potential associated with the spin-$1/2$ $J/\psi N$ channel can potentially form a loosely bound state, with an approximate binding energy of 0.1 MeV for the $c\bar{c}\textrm{-}\alpha$ system. Other configurations remain unbound relative to the $c\bar{c} + \alpha$ threshold. When the statistical uncertainties of the parameters were included, an upper limit of approximately 1.3 MeV was predicted for the binding energy, primarily associated with the spin-$\frac{1}{2}$ $J/\psi N$ component.

Furthermore, the spin-averaged WS-type potential for the $J/\psi\textrm{-}\alpha$ interaction was examined. This effective nuclear potential was derived through the single folding of the nucleon density within the $\alpha$-particle with the spin-averaged $J/\psi N$ interaction defined as:
\begin{equation}
J/\psi N^{\textrm{spin-ave.}}=\frac{2}{3}J/\psi N\left(^{4}S_{3/2}\right)+\frac{1}{3}J/\psi N\left(^{2}S_{1/2}\right). \label{eq:NJpsi-spin-ave}
\end{equation}
Studying the differences arising from these variations is especially useful in understanding their impact on the binding energy of the $ c\bar{c}\textrm{-} \alpha\alpha$ three-body ground state.

{\textit{$\alpha\alpha$ interaction- }
For the $\alpha\alpha$ potential, an Ali-Bodmer-like potential~\cite{Ali1966} was employed, composed of a central term expressed as the sum of two Gaussian functions:
\begin{equation}
	V_{\alpha\alpha}^{\left(l\right)}\left(r\right)=v_{R}^{\left(l\right)}\exp\left[-\left(r/1.53\right)^{2}\right]-30\:\exp\left[-\left(r/2.85\right)^{2}\right].
\end{equation}
where $v_{R}^{(l)}$ are the repulsive strengths, varying with the angular momentum quantum number $l$, to reproduce the experimental phase shifts. Specifically, $v_{R}^{(l)}$ takes values of 125 MeV for $l=0$, 20 MeV for $l=2$, and zero for higher partial waves $l>2$. When combined with a hard-sphere Coulomb interaction characterized by a Coulomb radius of $r_{Coul}=2.94$ fm, given by
\begin{equation}
	V_{Coul}\left(r\right)=Z^{2}e^{2}\times\begin{cases}
		\frac{1}{r_{Coul}}\left(\frac{3}{2}-\frac{r^{2}}{2r_{Coul}^{2}}\right); & r\leq r_{Coul}\\
		\frac{1}{r}; & r>r_{Coul}
	\end{cases} \label{eq:coulomb}
\end{equation}
this potential successfully reproduces the energy position of the $^{8}$Be s-wave resonance, which is essential for describing the continuum properties of low-lying $^{8}$Be states~\cite{Casal2016}.

In modern approaches, halo nuclear states have been addressed using effective field theory (EFT) ~\cite{Zhukov93,FREDERICO2012939}. These states are characterized by a nucleus composed of a tightly bound core and one or more weakly bound (valence) nucleons. In the first approximation, the core is treated as an explicit degree of freedom, and the EFT is formulated in terms of contact interactions between the valence nucleons and the core.  
The $\alpha\alpha$ interaction in an EFT framework for halo-like systems has been investigated by Higa, Hammer, and van Kolck~\cite{HIGA2008171}. Such interactions serve as critical inputs for systems involving more than two alpha particles in multi-body calculations, such as the triple-alpha reaction, which describes the formation of $^{12}$C through the Hoyle state.

At energies below the pion mass, the EFT for a two-alpha-particle system is constructed, in the first approximation, by representing each $\alpha$ particle with a scalar-isoscalar field. The strong interactions are encoded in a derivative expansion of local operators. Although a satisfactory description of the data is achieved, a significant amount of fine-tuning is required. However, the question of proper non-perturbative renormalization conditions for EFT incorporating Coulomb forces remains unresolved.  

\section{ Hyperspherical harmonics formalism  }
\label{sec:Three-body hyperspherical}
Since this approach has been extensively described in previous studies~\cite{Zhukov93, raynal1970, face,etminan2024prc}, only a brief overview of the formalism is provided here. This summary should be sufficient to define the relevant quantities and notation used in the subsequent sections.

The present study considers a three-body model comprising a $c\bar{c}$ pair and two $\alpha$ particles, labeled by indices \(i = 1, 2, 3\), with corresponding masses \(m_i\), position vectors \(r_i\), and momenta \(p_i\). These particles interact through pairwise potentials \(V_{ij}\). To describe the system, the Jacobi coordinates, which—apart from mass factors—are defined as the relative position between two particles (denoted by \(\boldsymbol{x}\)) and between their combined center of mass and the third particle (denoted by \(\boldsymbol{y}\)). These coordinates are illustrated in Fig.~\ref{fig:T_jacobi} and mathematically expressed as:
\begin{equation}
	\vec{x}_i = \sqrt{\frac{A_j A_k}{A_j + A_k}}\ \vec{r}_j - \vec{r}_k, \quad \vec{y}_i = \sqrt{ \frac{(A_j + A_k) A_i}{A_i + A_j + A_k}}\ \vec{r}_i - \frac{A_j \vec{r}_j + A_k \vec{r}_k}{A_j + A_k},
\end{equation}
where \(A_i ={m_i}/{m}\), and \(m\) is taken as 1 atomic mass unit, i.e., $m = m_N = 938.9 $ MeV.

The explicit forms of these coordinates, along with the three corresponding sets of hyperspherical coordinates \(\{\rho, \theta, \Omega_x, \Omega_y\}\), can be found in Refs.~\cite{Zhukov93, raynal1970, face}. Here, \(\rho^2 = x^2 + y^2\) represents the generalized radial coordinate, while the hyperangle \(\theta = \arctan(x/y) \in [0, \pi/2]\) characterizes the relative sizes of \(\boldsymbol{x}\) and \(\boldsymbol{y}\). The angle $\Omega_{x}\left(\Omega_{y}\right)$ describes the directions of $\boldsymbol{x}\left(\boldsymbol{y}\right)$. For brevity, all angular dependencies are encapsulated within the set $\varphi = (\theta, \Omega_x, \Omega_y)$.

In a three-body model consisting of three-particle or clusters such as $ c\bar{c}\textrm{-} \alpha\alpha$, the total three-body wave function $\Psi$ is expressed as $ \Psi=\psi^{\left(1\right)}+\psi^{\left(2\right)}+\psi^{\left(3\right)} $. 
The components $\psi^{(i)}$ are functions of three different sets of Jacobi coordinates. One such set is illustrated in Fig.~\ref{fig:T_jacobi}. These components satisfy the three coupled Faddeev equations,
\begin{equation}
	\left(T-E\right)\psi^{\left(i\right)}+V_{jk}\left(\psi^{\left(i\right)}+\psi^{\left(j\right)}+\psi^{\left(k\right)}\right)=0,
	\label{eq:faddeev_coupled-eq}
\end{equation}
where $T$ is the kinetic energy operator, $E$ the total energy, and $V_{jk}(r_{jk})$ the interaction potential between particles $j$ and $k$. The indices $i, j, k$ are cyclic permutations of $(1,2,3)$.

\begin{figure*}[hbt!]
	\centering
	\includegraphics[scale=0.15]{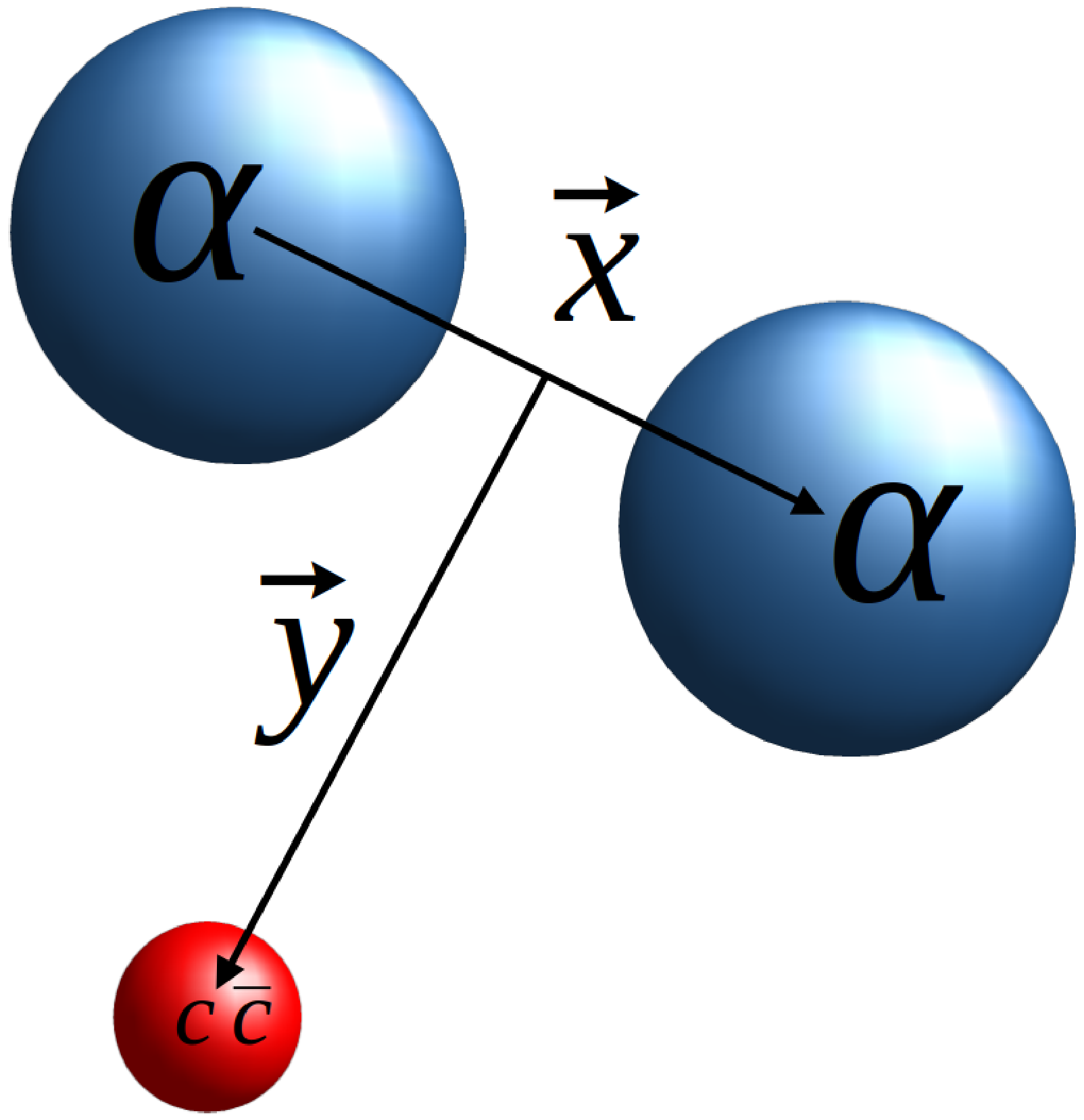}
	\caption{The Jacobi-T set, connects the two identical $\alpha$ particles by coordinate $\vec{x}$, is used to describe $ _{c\bar{c}}^{9}$Be as $ c\bar{c}\textrm{-} \alpha\alpha$. 
		The $\alpha$ particles are two identical bosons with spin $0^{+}$, so they couple to $S_{x} = 0$. }
	\label{fig:T_jacobi}
\end{figure*}

The Hamiltonian of a three-body system expressed in hyperspherical coordinates is
\begin{equation}
	\hat{H} = \hat{T}(\rho, \varphi) + \hat{V}(\rho, \varphi),\label{eq:Hamiltonian}
\end{equation}
where the potential operator $\hat{V}(\rho, \varphi)$ encompasses the sum of pairwise interactions. The kinetic energy operator $\hat{T}(\rho, \varphi)$, as defined in Ref.~\cite{lay2012}, takes the form
\begin{equation}
	\hat{T}(\rho, \varphi) = -\frac{\hbar^2}{2m} \left( \frac{\partial^2}{\partial \rho^2} + \frac{5}{\rho} \frac{\partial}{\partial \rho} - \frac{1}{\rho^2} \hat{K}^2(\varphi) \right),
\end{equation}
with $\hat{K}$ representing the hyperangular momentum operator, regarded as a generalized angular momentum operator. The mass parameter $m$ is chosen as the normalization mass which, for nuclear systems, is usually taken
as the atomic mass unit.

Solutions to the Schr\"odinger equation with this Hamiltonian Eq.~\eqref{eq:Hamiltonian} are expanded for each value of $\rho$, given the total angular momentum quantum number $j$, as
\begin{equation}
	\psi_{i \beta}^{j \mu}(\rho, \varphi) = R_{i \beta}(\rho) \mathcal{Y}_{\beta}^{j \mu}(\varphi),
\end{equation}
where the set of hyperangular functions $\mathcal{Y}_{\beta}^{j \mu}(\varphi)$ forms a complete basis. These functions are expanded in hyperspherical harmonics~\cite{Zhukov93,Casal2020,raynal1970}, and the hyperradial functions $R_{i \beta}(\rho)$ are introduced, with the index $i$ indicating hyperradial excitations. To facilitate solving the coupled differential equations, the hyperradial functions $\mathcal{R}_{\beta}^j(\rho)$ are expanded in an orthonormal discrete basis up to a maximum excitation $i_{\max}$~\cite{face}.

A set of quantum numbers $\beta = \{K, l_x, l_y, l, S_x, j_{ab}\}$ is used, where $l_x$ and $l_y$ denote the orbital angular momenta associated with the Jacobi coordinates $\vec{x}$ and $\vec{y}$, respectively. The total orbital angular momentum $l$ is given by $l_x + l_y$, and the total spin of the pair coupled via $\vec{x}$ is represented as $S_x$. The quantum number $j_{ab} = l + S_x$ and the total angular momentum is expressed as $j = j_{ab} + I$, with $I$ denoting the spin of the third particle.

The system’s wave function is expressed as
\begin{eqnarray}
	\psi^{j \mu}(\rho, \varphi) &=& \sum_{\beta} \sum_{i=0}^{i_{\max}} C_{i \beta}^j \ \psi_{i \beta}^{j \mu}(\rho, \varphi) \nonumber \\
	&=& \sum_{\beta} \left( \sum_{i=0}^{i_{\max}} C_{i \beta}^j R_{i \beta}(\rho) \right) \mathcal{Y}_{\beta}^{j \mu}(\varphi) \equiv \sum_{\beta} \mathcal{R}_{\beta}^j (\rho) \ \mathcal{Y}_{\beta}^{j \mu} (\varphi),
\end{eqnarray}
where the coefficients $C_{i \beta}^j$ are obtained via diagonalization of the Hamiltonian within the basis functions up to $i_{\max}$.
The hyperradial wave functions, denoted as \(\mathcal{R}_{\beta}(\rho)\), are obtained as solutions to the set of coupled differential equations and are expressed in the form
\begin{equation}
	\left(-\frac{\hbar^{2}}{2m} \left(\frac{d^{2}}{d\rho^{2}} - \frac{(K+3/2)(K+5/2)}{\rho^{2}}\right) - E \right) \mathcal{R}_{\beta}^{j}(\rho) + \sum_{\beta'} V_{\beta' \beta}^{j \mu}(\rho) \mathcal{R}_{\beta'}^{j}(\rho) = 0, \label{eq:Faddeev_coupled}
\end{equation}
where the term \(V_{\beta' \beta}^{j \mu}(\rho)\), representing the two-body potentials between pairs of particles \(V_{ij}\), is defined as
\begin{equation}
	V_{\beta' \beta}^{j \mu}(\rho) = \left\langle \mathcal{Y}_{\beta}^{j \mu}(\varphi) \middle| V_{12} + V_{13} + V_{23} \middle| \mathcal{Y}_{\beta'}^{j \mu}(\varphi) \right\rangle. \label{eq:V_ij}
\end{equation}
In this formulation, the hyperradial functions are solutions to these coupled equations, which incorporate the contributions from all pairwise interactions through the matrix elements \(V_{\beta' \beta}^{j \mu}(\rho)\).
		\section{Numerical results} 
		\label{result}	
The numerical results for the ground state central binding energy, denoted as \( B_3 \), and the nuclear matter radius, \( r_{\text{mat}} \), for the three-body $ c\bar{c}\textrm{-} \alpha\alpha$ system are presented and discussed herein. The coupled equations, as shown in Equations~\eqref{eq:Faddeev_coupled} and the angular integration in Eq.~\eqref{eq:V_ij} are solved following the prescriptions in the FaCE code~\cite{face}, with the two-body interactions described in Sec.~\ref{sec:Two-body-potentials}.

In the HH method, the final results are dependent on the maximum hypermomentum value, \( K_{\text{max}} \), due to the truncation of the total three-body wave function expansion in hypermomentum components. Consequently, an initial investigation into the convergence of results was performed as a function of \( K_{\text{max}} \) and the maximum number of hyperradial excitations, \( i_{\text{max}} \). For the \( c\bar{c}\textrm{-}\alpha \alpha \) state, convergence was found to be satisfactory with parameters \( K_{\text{max}} = 30 \) and \( i_{\text{max}} = 25 \).

The central binding energies of the two-body $ c\bar{c}\textrm{-} \alpha\alpha$ systems, $B_{c\bar{c}\textrm{-}\alpha}$, are obtained in~\cite{etminan2025alphaCcbar}.
As mentioned in the previous section, none of the \(c\bar{c}\textrm{-}\alpha \) two-body potentials, with parameters listed in Table~\ref{tab:Charm-alpha-para}, form bound states. 
The only exception is the potential based on the spin-\(\frac{1}{2}\) $J/\psi N$ interaction, which results in a loosely bound state with approximately 0.1 MeV of central binding energy relative to the \( c\bar{c} + \alpha \) threshold.

The ground state central binding energies and nuclear matter radii of the \(c\bar{c}\textrm{-}\alpha \alpha \) nuclei are summarized in Table~\ref{tab:Charm-alpha-para} for the \(\ J/\psi N(^{4}S_{3/2}) \), \(\ J/\psi N(^{2}S_{1/2}) \), \(\ \eta_c N(^{2}S_{1/2}) \), and the spin-averaged \( J/\psi N \) interaction at Euclidean time \( t/a=14 \). 
Since the $J/\psi$–nucleon interaction is spin-dependent, the dependence of the ground state characteristics of the $ c\bar{c}\textrm{-} \alpha\alpha$ system on the spin of the $J/\psi$–nucleon interaction is investigated. In Table~\ref{tab:Charm-alpha-para}, the first column which is remarked by "Based interaction" displays the $c\bar{c}$-N interaction selected for extracting the $J/\psi\textrm{-}\alpha$ potential, which is derived by folding the $J/\psi$–N interaction.
As expected, all four derived $J/\psi\textrm{-}\alpha$ potentials are found to be attractive.

The maximum binding energy for the $ c\bar{c}\textrm{-} \alpha\alpha$ system is approximately $ 1.71(1.60) $ MeV, corresponding to the \( J/\psi N(^{2}S_{1/2}) \) interaction, while the minimum is about $ 0.56(30) $ MeV, associated with the \( \eta_c N(^{2}S_{1/2}) \) interaction.
Moreover, the $J/\psi\textrm{-}\alpha$ potential is obtained by folding the spin-averaged \( J/\psi N \) interaction, as defined by Eq.~\eqref{eq:NJpsi-spin-ave}. The corresponding fitting parameters are provided in the last row of Table~\ref{tab:Charm-alpha-para}.  
In this case, the three-body binding energy is determined to be $1.31(1.10)$ MeV, a value that lies approximately between those corresponding to the spin components \(\ J/\psi N(^{4}S_{3/2}) \) and \(\ J/\psi N(^{2}S_{1/2}) \).

Furthermore, results obtained using the isospin-averaged hadron masses from \((2+1)\)-flavor lattice QCD simulations~\cite{LyuPLB2025} are also presented in Table~\ref{tab:Charm-alpha-para} within square brackets. These masses are slightly higher than the experimental values. As illustrated in the table, the \( B_3 \) values derived from lattice masses are marginally larger than those obtained from experimental masses. This trend is attributed to the increased masses, which decrease the repulsive kinetic energy contribution, thereby leading to an increase in binding energies~\cite{Garcilazo2019}.

The matter radius for an \(A\)-nucleon system is calculated using the relation
\begin{equation}
	r_{\text{mat}} = \sqrt{\left\langle r^2 \right\rangle} = \sqrt{\frac{1}{A} \left\langle \sum_{i=1}^{A} r_i^2 \right\rangle}, \label{eq:r_mat}
\end{equation}
where \(\boldsymbol{r}_i\) denotes the position of each nucleon relative to the system’s center of mass. To compute the root-mean-square (rms) matter radius of the $ c\bar{c}\textrm{-} \alpha\alpha$ state, the rms matter radius of the \(\alpha\) particle—taken as 1.47 fm~\cite{ANGELI201369}—and the strong interaction radius of the \( J/\psi \), 0.25 fm~\cite{POVH1990653}, are employed. The resulting nuclear matter radii of the $ c\bar{c}\textrm{-} \alpha\alpha$ bound states also presents in Table~\ref{tab:Charm-alpha-para}.	
\begin{table}[hbt!]
	\centering
	\caption{
		The three fitting parameters, \( U_0 \), \( R \), and \( t \), of the \( U_{c\bar{c}\textrm{-}\alpha}\left(r\right) \) potential in Eq.~\eqref{eq:ws-fit} are specified. 
		$B_{c\bar{c}\textrm{-}\alpha}$ is the central binding energy of the two-body \(c\bar{c}\textrm{-}\alpha \) system.
		The central binding energy \( B_3 \) and the nuclear matter radius \( r_{\text{mat}} \) of the ground state of the three-body 
		$ c\bar{c}\textrm{-} \alpha\alpha$ system are calculated. These calculations are performed using the experimental masses, specifically
		 \( m_\alpha = 3727.38 \) MeV/\( c^2 \), \( m_{J/\psi} = 3096.9 \) MeV/\( c^2 \), and \( m_{\eta_c} = 2984.1 \) MeV/\( c^2 \). Additionally, the results corresponding to the \( c\bar{c} \) mass value obtained through lattice simulations~\cite{LyuPLB2025} are provided within square brackets. The numbers in parentheses are statistical errors.
		\label{tab:Charm-alpha-para}}	
	\begin{tabular}{cccccccc}
		\hline
		\hline 
	 Based interaction & $ U_{0} $ (MeV)& $ R $ (fm)& $ t $ (fm)&$B_{c\bar{c}\textrm{-}\alpha}$ (MeV) && $B_3$ (MeV) & $r_{mat}$ (fm) \\
		\hline
 spin 3/2 $J/\psi N $       & $  8.95(3.35)$ & $ 1.684(5) $  & $ 0.367(10) $ & -         &&$ 1.13(1.00)[1.13]$ & $ 2.75(41)$ \\
 spin 1/2 $J/\psi N $       & $ 10.79(5.25)$ & $ 1.668(11) $ & $ 0.366(20) $ & 0.1       &&$ 1.71(1.60)[1.71]$ & $ 2.60(51)$ \\
 spin 1/2 $\eta_{c} N$      & $ 7.12(1.39) $ & $ 1.677(6) $  & $ 0.360(5)  $ & -         &&$ 0.56(30)[0.57]$   & $ 3.05(49)$ \\
 spin-averaged $J/\psi N$   & $ 9.56(3.97) $ & $ 1.678(6) $  & $ 0.367(13) $ & -         &&$ 1.31(1.10)[1.31]$ & $ 2.70(35)$\\ 
		\hline
		\hline 	
	\end{tabular}
\end{table}

The characteristic features of Borromean bound states are characterized by weak binding and large spatial extension, as demonstrated in Ref.~\cite{PhysRevLett.76.3903}. The spatial extension of the system can be easily estimated by the root mean square radius, as provided in Eq.~\eqref{eq:r_mat}.  
To provide a more explicit demonstration of the Borromean behavior of the \( _{c\bar{c}}^{9}\)Be system, its root mean square radius is compared with that of the ordinary Borromean \(^{9}\)Be nucleus. The \(^{9}\)Be nucleus is described as a three-body system, consisting of two \(\alpha\) particles and one neutron. This configuration exhibits a Borromean structure, as neither of the binary subsystems—\(^{5}\)He nor \(^{8}\)Be—are bound. The \(3/2^{-}\) ground state of \(^{9}\)Be, however, is weakly bound, with a binding energy of 1.5736 MeV below the \(\alpha+\alpha+n\) threshold~\cite{TILLEY2004155}.  

For the \(^{9}\)Be ground state, the experimental charge radius is measured as \(2.519\pm0.012\) fm~\cite{ANGELI201369}, while the matter radius, derived from Ref.~\cite{TANIHATA1988592}, is \(2.38\pm0.01\) fm. The latter was obtained through Glauber-model calculations based on interaction cross sections at high energies. By comparing the nuclear matter radius \( r_{\text{mat}} \) of the $ c\bar{c}\textrm{-} \alpha\alpha$ system (as listed in Table~\ref{tab:Charm-alpha-para}) with the matter and charge radii of the \(^{9}\)Be nucleus, it is observed that the \( r_{\text{mat}} \) values for the
 $ c\bar{c}\textrm{-} \alpha\alpha$ system comparatively exceed those of the \(^{9}\)Be nucleus. This observation implicitly confirms the Borromean structure of the $ c\bar{c}\textrm{-} \alpha\alpha$ system.

Although significant experimental challenges would be posed by the creation and detection of such systems,  
potential observations of these $c\bar{c}\textrm{-}\alpha\alpha$ states could be made through hadron-beam experiments at facilities such as Jefferson Lab, J-PARC, and FAIR, or through relativistic heavy-ion collisions at RHIC and the LHC.  
	
\section{Summary and outlook\label{sec:Summary-and-conclusions}}
The possible existence of a bound state of $ _{c\bar{c}}^{9}$Be, a charmonium-nucleus system, has been investigated. The analysis was conducted within a three-cluster model, in which its binary subsystems are represented as $ c\bar{c}\textrm{+}\alpha $ and $\alpha+\alpha$. A method based on hyperspherical harmonics was employed to enable a convenient and accurate description of this three-cluster configuration.

Effective nuclear potentials for the $ c\bar{c}\textrm{-}\alpha $ interaction were estimated through single folding of the nucleon density within the $\alpha$-particle. These potentials were derived from state-of-the-art HAL QCD calculations, which provided interactions for 	the spin 3/2 $J/\psi N $, spin 1/2 $J/\psi N $, spin 1/2 $\eta_{c}N$ and spin-averaged $J/\psi N$ interactions, all obtained with nearly physical pion masses. Additionally, the Coulomb interaction was incorporated into the calculations.

The nucleus $ _{c\bar{c}}^{9}$Be was modeled as a three-body system composed of two $\alpha$-particles and one charmonium. Numerical results indicated that the $ c\bar{c}\textrm{-} \alpha\alpha$ system could potentially form bound states. The maximum central binding energy was found to be approximately $ 1.71(1.60) $ MeV, based on the spin-1/2 $J/\psi N $ interaction, while a minimum value of around $ 0.56(30) $ MeV was obtained using the spin-1/2 $\eta_{c}N$ interaction. For the $\alpha \alpha$ interaction, the widely used Ali-Bodmer double Gaussian potential was employed, which has been extensively applied in calculations of nuclear binding energies within the cluster model framework.

It was suggested that $ _{c\bar{c}}^{9}$Be could be a Borromean nucleus, where the three-body system is bound, albeit loosely, despite none of the binary subsystems having bound states. These are genuinely non-mean-field systems, and a three-body treatment is therefore the natural approach, utilizing binary interaction potentials derived from lattice QCD studies. It has been demonstrated that such calculations are feasible with current techniques. Furthermore, ab initio methods that address Borromean nuclei should incorporate three-cluster dynamics to properly account for the intrinsic three-body character of the systems.

The intriguing physics of heavy-flavor nuclei should eventually be explored in practical experiments. To achieve this, the production of heavy particles, such as charmonium, must first be considered in a controlled environment. Subsequently, the formation of the $_{c\bar{c}}^{9}$Be system, achieved through interactions of charmoniums with $^{8}$Be nuclei, should be guaranteed. Finally, the structure of this charmonium-nucleus system must be investigated.

In experiments, heavy hadrons and heavy-flavor nuclei are typically produced via hadron beams, relativistic heavy-ion collisions, photoproduction, electron-nucleus scattering, and neutrino-nucleus reactions~\cite{HOSAKA201788}. 
On the experimental side, a major challenge lies in identifying appropriate kinematical conditions under which these hadrons are produced essentially at rest or with minimal momentum relative to the nucleus. Since the effects of the nuclear medium are governed by low-energy interactions~\cite{KREIN2011136}, such kinematics may be attainable through subtle selection of beam and target nuclei~\cite{KreinPRC2011}. 

And finally, since the formed Borromean, halo or cluster states frequently emerge near the boundaries of nuclear stability, they are influencing ongoing and proposed experiments at the Facility for Rare Isotope Beams (FRIB) and other radioactive beam facilities worldwide~\cite{bazin2023perspectives}. For instance, the Active Target Time Projection Chamber (AT-TPC)~\cite{BRADT201765} is ideally suited for probing the cluster structure of low-energy resonances, as it enables resonant scattering reactions to be performed by scanning various beam energies in a single experiment.

Although the calculations are performed within the Faddeev-equation framework, which is regarded as the most accurate approach in few-body physics, a detailed, full-scale many-body calculation, such as no-core shell model~\cite{BARRETT2013131} or Gaussian expansion method~\cite{HIYAMA2003223}, employing realistic nucleon-nucleon and meson-nucleon interactions, is required for an accurate comparison with future experimental data.

In order to further enhance the results, it is currently underway by the HAL QCD Collaboration to perform calculations at the physical point using \((2+1)\)-flavor lattice QCD configurations, which have been generated by them~\cite{ScalePhysRevD.110.094502}.
It is likely that additional structureless hyperradial three-body forces may be needed to complement binary interactions, thereby enabling theoretical predictions to be aligned with future experimental data.  
The validity of these cluster assumptions can be best justified by comparing calculation results with future measurements.

\bibliography{Refs.bib}
\end{document}